\documentstyle[12pt,fleqn,epsfig]{article}

\newcommand{\be}{\begin{equation}}
\newcommand{\ee}{\end{equation}}
\newcommand{\bear}{\be\begin{array}}
\newcommand{\eear}{\end{array}\ee}
\newcommand{\bea}{\begin{eqnarray}}
\newcommand{\eea}{\end{eqnarray}}

\catcode`@=11
\newtoks\@stequation
\def\subequations{\refstepcounter{equation}%
\edef\@savedequation{\the\c@equation}%
 \@stequation=\expandafter{\theequation}
 \edef\@savedtheequation{\the\@stequation}
 \edef\theequation{\theequation}%
 \setcounter{equation}{0}%
 \def\theequation{\theequation\alph{equation}}}
\def\endsubequations{\setcounter{equation}{\@savedequation}%
  \@stequation=\expandafter{\@savedtheequation}%
  \edef\theequation{\the\@stequation}\global\@ignoretrue
\noindent}
\catcode`@=12

\begin{document}

\begin{flushright}
UCLA/04/TEP/53\\
December 2004
\end{flushright}

\baselineskip 18pt

\begin{center}
{\Large {\bf  Prospect for relic neutrino searches}}\footnote{Talk given at the Nobel Symposium on Neutrino Physics, Enkoping, Sweden, Augus 19-24, 2004.}
 \vskip5mm Graciela B. Gelmini$^{1,}$\footnote{Email address:
 gelmini@physics.ucla.edu},
 \vskip3mm \mbox{}%
$^{1}$ Physics and Astronomy Department, University of California
 Los Angeles (UCLA),
405 Hilgard Ave., Los Angeles CA 90095, USA

\bigskip

\begin{abstract}

Neutrinos from the Big Bang are theoretically expected to be the
most abundant particles in the Universe after the photons of
the Cosmic Microwave Background (CMB). Unlike the relic photons,
relic neutrinos have not so far  been observed. The Cosmic Neutrino
Background (C$\nu$B) is the oldest relic from the Big Bang,
 produced a few seconds
after the Bang itself.  Due to their impact in cosmology,
 relic neutrinos may be
revealed indireclty in the near future through cosmological observations.
In this talk we concentrate on other proposals, made in the last 30 years,
to try to detect the C$\nu$B directly, either in
laboratory searches (through tiny
accelerations they produce on macroscopic targets) or through
 astrophysical observations
(looking  for absorption dips in the flux of
Ultra-High Energy (UHE) neutrinos, due to the annihilation of these neutrinos
with  relic neutrinos at the Z-resonance).

We concentrate mainly on the first possibility.
We show that, given present bounds on neutrino masses, lepton number
in the Universe and gravitational clustering of neutrinos, all expected
laboratory effects of relic neutrinos are  far from observability,
awaiting  future technological advances to reach the necessary
sensitivity. The problem for astrophysical searches is
that  sources of UHE neutrinos at the extreme energies required may not
exist. If they do exist, we could reveal the existence, and possibly the
mass spectrum, of relic neutrinos, with detectors of UHE neutrinos (such
as ANITA,  Auger,  EUSO,  OWL, RICE and SalSA).

\bigskip
\noindent PACS numbers: 14.60.Lm; 14.60.Pq; 95.35.+d; 95.85.Ry

\end{abstract}
\end{center}
\newpage

\section*{1. Introduction}

Neutrinos from the Big Bang are theoretically expected to be the
most abundant particles in the Universe after the photons of
the Cosmic Microwave Background (CMB). The Cosmic Neutrino Background
C$\nu$B can contain the three active neutrinos of the Standard Model of 
Elementary Particles (SM), which are Dirac or Majorana particles with masses 
between about 0.01 and  1 eV, one or more sterile neutrinos
 (present in trivial extensions
 of the SM) and possibly light bosons coupled to neutrinos.
Unlike the CMB,
the C$\nu$B has not been yet observed. Its  detection would provide
insight into early moments of our Universe, from before Big-Bang the
 Nucleosynthesis 
(BBN) until 
now. In fact, the C$\nu$B is the oldest relic from the Big Bang,
 produced a few seconds
after the Bang itself. Thus, it impacts cosmology from the BBN
 (which finished about 20 minutes later), to the emission
the CMB (380 kyr later), to the formation the Large Scale Structure of the
Universe (a Gyr later). Due to this impact, relic neutrinos may be
revealed indireclty in the near future through cosmological 
observations~\cite{tegmark}.

In this talk we concentrate on other proposals, made in the last 30 years,
to try to detect the C$\nu$B, either in
laboratory searches or through astrophysical observations. We will
concentrate in the first possibility and mention the second briefly.

In laboratory experiments cosmic neutrinos could be revealed through the tiny
accelerations they produce on macroscopic targets, accelerations which are
quadratic or linear in the Fermi coupling constant. Forces quadratic in
the Fermi constant are for sure present and are largest for Dirac
neutrinos. Torques linear in the Fermi coupling constant could be present
only if there is a net Lepton number in the background,  i.e. a
difference between relic neutrinos and antineutrinos. If present, this
effect has a comparable  magnitude for both Majorana and Dirac neutrinos.
We show that, given present bounds on neutrino masses, Lepton number
in the Universe and gravitational clustering of neutrinos, all expected
laboratory effects of relic neutrinos are  far from observability,
awaiting  future technological advances to reach the necessary
sensitivity.

Astrophysical searches would look for absorption dips in the flux of
Ultra-High Energy neutrinos, due to the annihilation of these neutrinos
with  relic neutrinos at the Z-resonance. The problem with this idea is
that  sources of UHE neutrinos at the extreme energies required
 ($10^{22}$ eV) may not
exist. If they do exist, we could reveal the existence, and possibly the
mass spectrum, of relic neutrinos, with detectors of UHE neutrinos, such
as ANITA,  Auger,  EUSO, OWL, Rice and SalSA.

\section*{2. The standard  relic neutrino background }

The standard relic neutrino background is assumed to consist of
 the three active neutrinos of the SM,
with relic abundances dictated purely by the interactions present
 in the SM, and a negligible Lepton number in the Universe. However
 we know that there must be neutrino physics beyond the SM, because
we know now experimentally that neutrinos have masses, which they 
do not in the SM. In fact the 
solar mass-square difference $\Delta m^2_{12} \simeq 8.1 \times 10^{-5}$ eV$^2$
and the atmospheric mass difference
$\Delta m^2_{23} \simeq 2.2 \times 10^{-3}$ eV$^2$~\cite{deltam} 
emply that there are at least three neutrino mass eigenstates. 
Since the larger mass entering into a mass-square 
difference must be larger than or equal to 
 $\sqrt{ \Delta m^2}$, we know that two of the three
 masses must be larger than 0.9 $\times 10^{-2}$ eV. Cosmological bounds
 dictate that all active neutrino masses are smaller than about 1
 eV~\cite{eVmass}.

The thermal history of neutrinos starts at temperatures  $T>>$ MeV,
 at which the three active neutrinos $\nu_\alpha$ of the SM ($\alpha$
 stands for e, $\mu$ or $\tau$) 
 were in equilibrium, its reaction rate being larger than
 the expansion rate of the
 Universe, $\Gamma_\nu >> H$. Thus neutrinos had an equilibrium distribution
\begin{equation}
f_{\nu_\alpha}(p)=\left[\exp\left ( \frac{E-\mu_\alpha}{T}\right
)+1\right]^{-1}
\label{fnu}
\end{equation}
where $E\simeq p$, since $m_\alpha<< $MeV and 
$\mu_\alpha$'s are the chemical potentials. The standard
 assumption is that $\mu_\alpha=0$.
At $T \simeq$ MeV,   the neutrino interaction rate,
 $\Gamma_\nu=\langle\sigma_\nu n_\nu\rangle$, falls
below the expansion rate, $H=\sqrt{8\pi\rho/3M_P^2}$, and
 neutrinos decouple at a temperature
 between  2 and 3 MeV, depending on the flavour. But even
 after neutrinos are decoupled, while they are relativistic
 they maintain  the  equilibrium distributions $f_{\nu_\alpha}(p)$.
  Just after neutrinos decouple,
 at $T \simeq m_e= 0.5$ MeV, $e^\pm$  pairs annihilate and 
transfer their entropy into
photons, increasing their temperature $T$ with respect to the
 temperature of neutrinos, which becomes 
 $T_\nu = \left( 4/11 \right)^{1/3}T$.  Therefore now
 $T_\nu= 1.9 ^oK= 1.7~10^{-4}$eV, which means that at least two of
 the active neutrinos in the C$\nu$B  (with masses above $10^{-2}$eV) are 
 non-relativistic. The Dirac or Majorana
 nature of neutrinos becomes important for non -relativistic neutrinos.
 
 From this history we obtain the usual expressions for the number density
\begin{equation}
n_{\nu} = n_{\nu^c}= \frac{3}{22}\;n_\gamma 
= \frac{3\zeta(3)}{11\pi^2}\;T^3=  56/{\rm cm^3}~,
 \label{nnu}
\end{equation}
the relativistic energy density (for $T_\nu > m_\nu$)
\begin{equation}
\rho_\nu + \rho_{\bar{\nu}}
= \frac{7}{8} \left(\frac{T_\nu}{T}\right)^4 \rho_\gamma=
\frac{7\pi^2}{120}
\left(\frac{4}{11}\right)^{4/3}\;T^4~,
 \label{rhoR}
\end{equation}
and the non-relativistic energy density
 (for $T_\nu <  m_\nu$) $\rho_\nu =  m_\nu n_\nu$ (using
$\Omega$, the density in units of the critical density, and $h$,
the reduced Hubble constant), i.e. 
\begin{equation}
 \Omega_\nu h^2 =\sum_\alpha m_\alpha / ~94 {\rm eV}~, 
 \label{rhoNR}
\end{equation}
which, with only the interactions present in the SM, are
 the same for Majorana or Dirac neutrinos.
The reason is that although Dirac neutrinos have four possible
 states and Majorana only two,
 the two additional states of Dirac neutrinos are not
 populated in the early Universe, because of
how small neutrino masses are.

The number of relativistic neutrinos species, $N_\nu$, is
 used to parametrize  $(\rho_{\rm relativistic} - \rho_\gamma)$
 in terms of the
present density of one relativistic standard neutrino species (computed 
in the limit of instantaneous decoupling), i.e.
\begin{equation}
\rho_{\rm relativistic} = \rho_\gamma + \rho_\nu + \rho_x =
\left[ 1 + \frac{7}{8} \left( \frac{4}{11}
\right)^{4/3} \, N_{\nu} \right] \, \rho_\gamma~.
 \label{Nnu}
\end{equation}
Lower bounds on $N_\nu$ larger than zero have been obtained,
 both during BBN as well as from CMB measurements. The BBN
 two-$\sigma$ bound $1.6 \leq N_\nu \leq 3.2$~\cite{barger1}
 constitutes, in fact,  a detection of relic neutrinos, since during BBN
 at least $\nu_e,\bar\nu_e$ are needed (for the weak
 interactions of p, n). However, the two-$\sigma$ bound obtained by WMAP
 from measurements of the CMB
 anisotropy,  $0.9\leq N_\nu \leq 8.3$~\cite{barger1}, measures only
 the relativistic energy density in the Universe at the time of CMB emission
(380 kyr after the Bang), which may not necessarily
 consist of relic neutrinos.

\section*{3. Non-Standard neutrino backgrounds}

A non-standard relic neutrino background can have
less or more neutrinos that the standard background. 
In inflationary models, the beginning of the radiation dominated era of
the Universe results from the decay of coherent oscillations of a
scalar field, and the subsequent thermalization of the decay products
into a thermal bath with the so called ``reheating temperature", $T_R$.
This temperature may have been as low as 0.7 MeV~\cite{kohri} (a very
recent analysis strengthens this bound to $\sim$ 4
MeV~\cite{hannestad}). It is well known that a low reheating 
temperature inhibits the production of particles which would have become
non-relativistic or decoupled  at $T$ above or close to 
$T_R$~\cite{giudice1, giudice2}. The
final number density of active neutrinos starts departing from the
standard number for $T_R \simeq 8$~MeV, stays within 10\% of it for
$T_R > 5$~MeV, and
for $T_R = 1$~MeV the number of 
$\nu_{\mu,\tau}$ is about 2.7\% of the standard number.
 We may even have no relic neutrinos left
in extreme models in which neutrinos  annihilate into light
 boson at late times~\cite{nudecay}.

A non-standard relic neutrino background with a neutrino 
asymmetry would have more neutrinos
than a standard background (in which the neutrino chemical
 potentials are assumed to be zero).
While charge neutrality requires the asymmetry in charged
 leptons to be the same
as that in protons, for which ($n_B-n_{\bar B})/n_\gamma \simeq O(10^{-10})$, 
no such requirement limits the asymmetry in neutrinos. With 
a relic neutrino asymmetry, the number of neutrinos and
 antineutrinos of the same flavour is different,
$n_{\nu_\alpha} \not= n_{\bar\nu_\alpha}$. The relic neutrino
 energy density always increases
with a neutrino asymmetry,
\begin{equation}
N_{\nu} =  3 + \frac{15}{7} \sum_\alpha \left [ 2 \left
(\frac{ \xi_\alpha}{\pi}\right )^2 + \left
(\frac{ \xi_\alpha }{\pi}\right )^4 \right ] = 3 +\sum_\alpha 
0.22\bigg(2~\xi_\alpha^2+0.10 ~\xi_\alpha^4 \bigg)~.
 \label{Nwithxi}
\end{equation}
We see here that for any non-zero value of the dimensionless chemical potential
$\xi_\alpha\equiv\mu_{\nu_\alpha}/T$ (chosen as parameter
 because it is constant while
the expansion of the Universe is adiabatic),  $N_{\nu}$ is larger
 that for a zero value, even if
for any value of $\xi$ smaller than 1 the increase is very small. For example, 
$ \Delta N_\nu = 4 \times 10^{-3}$ for $|\xi|= 0.1$ and
 $\Delta N_\nu \simeq 1$ for $|\xi|= 1.5$.

The net lepton number,
\begin{equation}
L_\alpha\equiv {\frac { n_{\nu_{\alpha}}-n_{\bar\nu_{\alpha}}} {n_\gamma}}=
{\pi^2 \over 12 \zeta(3)}\bigg(\xi_\alpha+{\xi_\alpha^3 \over \pi^2}\bigg)
\bigg({T_\nu \over T_\gamma} \bigg)^3=
0.25\bigg(\xi_\alpha+0.10~ \xi_\alpha^3 \bigg)
 \label{Lwithxi}
\end{equation}
can be sizeable even with values of $\xi$ somewhat smaller than 1. For example
$(n_{\nu_{\alpha}}-n_{\bar\nu_{\alpha}})\simeq 10/{ \rm cm}^3$ for
 $|\xi|= 0.1$ and
$(n_{\nu_{\alpha}}-n_{\bar\nu_{\alpha}})\simeq 190/{\rm cm}^3$ for
 $|\xi|= 1.5$.

So $|\xi| \geq 0.1$ produce small $\nu$-density increases but
 significant $\nu$-asymmetries.
This is important for what follows, because the most conservative
 upper bound on $\xi$ is about 0.1.

 In the absence of significant extra contributions to the radiation density
 of the
 Universe during BBN (except for that implied by the neutrino asymmetries),
 bounds from BBN alone allow for larger chemical potentials for $\nu_\mu$ and
$\nu_\tau$ than for 
$\nu_e$: $|\xi_e| < 0.2$, $|\xi_{\mu , \tau}| < 2.6$. The reason is that
 the effects of $\xi_e$ and  of $\xi_{\mu , \tau}$ compensate each other, but
 while the neutron to proton 
 ratio increases with $ \sqrt{\rho_{rad}}$ (which increases with 
 $\xi_{\mu , \tau}$ of any sign), it  
is proportional to exp$(-\xi_e)$), thus it decreases  faster with
positive $\xi_e$~\cite{kang}. However, due to the
 large mixings between neutrinos,
fast neutrino oscillations equilibrate all $\xi$, so the bound
 on all neutrino asymmetries must
be equal to the smallest upper bound imposed by BBN,
 $|\xi_{e , \mu , \tau}| <0.1$~\cite{lunardini}. These
 oscillations could be suppressed~\cite{babu} if neutrinos are coupled
 to light bosons
(Majorons~\cite{majorons}), whose interactions
 produce a large effective potential for neutrinos in the early
 Universe, and, if so,  the previously mentioned bounds would apply.
Even without the suppression of oscillations, the smallest
 upper bound imposed by BBN can be 
larger than 0.1, if an independent source of radiation is
 present during BBN, so that $\Delta N_\nu = N_\nu -3$ and the neutrino
 chemical potentials are independent parameters.
 Then, BBN combined with CMB data
(provided by WMAP) require $-0.1 \leq \xi_e \leq 0.3$
 for $-2 \leq \Delta N_\nu \leq 5$~\cite{barger2}.

Thus, in the following we will take as conservative upper bounds 
$\xi_{e , \mu , \tau}\simeq 0.1$ which implies 
$(n_\nu- n_{\bar\nu}) \simeq  10/cm^3$ and
 $(n_\nu+ n_{\bar\nu}) \simeq  112/cm^3$, or,
as an extreme upper bound $\xi_{ \mu , \tau}\simeq 3$, i.e. 
$L\simeq 2.5$, $(n_\nu- n_{\bar\nu}) \simeq
  n_\nu \simeq 1,050/cm^3$.

\section*{4.  Gravitational clustering of relic neutrinos}

Gravitational clustering of neutrinos in our galaxy or
 galaxy cluster may enhance the relic neutrino density
 making it easier to detect the C$\nu$B on Earth. 
 Already in  1979, Tremaine and Gunn~\cite{tremaine-gunn}
 produced a kinematical constraint, 
which shows  that neutrinos as light as we now know they
 are, would not significantly cluster.  Light 
neutrinos with masses $m_\nu <$ eV can be 
gravitationally bound only to the largest structures, large
clusters of galaxies. We can see this  using simple velocity arguments.
 Only cosmic neutrinos with velocities smaller
 than the escape velocity of a given structure can be bound to it.
The escape velocity from a large galaxy 
like ours is about 600 km/s and from a large cluster of
 galaxies is about 2,000 km/s.
Considering that the average velocity modulus  of  non-relativistic neutrinos
 of mass $m$ and  temperature  $T_\nu$ is  
(using Maxwell-Bolztman distribution)
 $\langle|\vec\beta_\nu|\rangle$=$\sqrt{8kT_\nu/
 \pi m}$=$\sqrt{4.3 ~10^{-4}{\rm eV}/m}$
 (namely $\langle|\vec v_\nu|\rangle$= 6,200 km/s for $m=1$eV, 
 and $\langle|\vec v_\nu|\rangle$= 19,600 km/s for $m=0.1$eV), 
it is obvious that only about a third of 1eV mass neutrinos, and  a very small
fraction of lighter neutrinos,  could be gravitationaly bound
 to large clusters at present.
Fermi degenerate neutrinos (those with $\xi>1$) may have even larger average
 velocities depending on their
chemical potential, 
but the conclusions remain the same. For  $\xi>>1$, we have
 $\langle|\vec\beta_\nu|\rangle$=$\sqrt{6 \xi T_\nu/ 5 m}
  \simeq \sqrt{ \xi 1.68~10^{-4}{\rm eV}/m}$ (namely 
  $\langle|\vec v_\nu|\rangle$= $\sqrt{\xi}$ 12,300 km/s
 for $m=0.1$eV), and both expressions 
  coincide for $\xi= 2.5$.
In all cases the amount of neutrinos in the tail of the
 velocity distribution with
velocities 
smaller that 600km/s, which would be gravitationally bound
 to galaxies, is much smaller.

More recently, in 2002, Singh and Ma~\cite{ma} studied the
 clustering of neutrinos in cold dark
matter halos. They found that neutrino overdensities decrease with 
cluster halo mass and distance to the center
 (and we are in the periphery of the Virgo supercluster), so
 overdensities close to Earth could be at most of  $O(1)$
 for neutrino masses close to 1 eV, and for $m_\nu \leq 0.1$ eV
 neutrino clustering is insignificant.

\section*{5. Prospects for laboratory searches:  effects of O($G_F^2$)}

 Given the characteristic relic neutrino energy 
 $\langle E_{\nu_\alpha}\rangle \simeq T_\nu \simeq  10^{-4}$ eV, the
relic $\nu$-nucleon cross sections are very small. For Dirac neutrinos,
\begin{equation}
\sigma_{\nu-N} \approx \left\{\begin{array}{ll}
G_F^2 m_\nu^2/\pi \simeq 10^{-56} \left(m_\nu / {\rm   eV} \right)^2
{\rm cm^2} & {\mbox{for (NR- Dirac)}} \\
G_F^2 E_\nu^2/\pi \simeq 5 \times 10^{-63} 
{\rm cm}^2 & {\mbox{for (R)}}
\end{array} \right.~,
 \label{sigma}
\end{equation}
where R stands for relativistic, NR for non relativistic
 and $G_F$ is the Fermi coupling constant. Majorana and Dirac
 neutrinos are indistinguishable while relativistic.
 For NR Majorana neutrinos the cross sections are even smaller:
since only  the $\gamma_\mu \gamma_5$ weak interaction coupling
 remains, a factor  $ \beta_\nu^2$ appears, where $\beta$ is the
 neutrino velocity. With $(n_\nu + n_{\bar\nu}) \simeq 100$ cm$^{-3}$,
 incoherent scattering off nucleons  leads to 
rates smaller than $ 10^{-6}$yr$^{-1}$ per  kiloton, for  the most favourable
 case NR Dirac neutrinos with eV mass.
 
Nuclear coherence enhancement factors, of order  $A^2 \simeq 10^4$
 (with $A$ the atomic number), do not help much. But coherence over
 the relic $\nu$ wavelength,  
$\lambda_\nu=2\pi\hbar/4T_\nu$ $\approx$ 2.4 mm (or
  1.2 mm$\times$eV/m$_\nu$ for 
clustered neutrinos),  makes an enormous
 difference since a
volume $\lambda_\nu^3$ contains more than $10^{20}$ nuclei. Since 
destructive interference occurs if target size is larger than
$\lambda_\nu$ the largest enhancement is obtained with a 
 material less than half filled with grains of size
  $ \lambda_\nu =$~\cite{zeldovich, smith}.
  
  Even with this sizeable cross section, the net momentum
 imparted by relic neutrinos on a target on Earth would be
 zero on average, if the C$\nu$B  would be on average at rest with
  respect to the Earth. But this is clearly not so. A reasonable
 guess is that 
  the C$\nu$B is at rest with the CMB, and the Sun's motion
 with respect to the CMB 
  (derived from COBE-DMR dipole anisotropy) is
 $v_{sun}=369.0\pm2.5$ km/sec, i.e. the speed of the Earth with respect to
 the C$\nu$B is
$\beta_{earth}=1.231\times 10^{-3}$. 
  
A momentum of the order of the neutrino momentum,
 $\Delta\vec p \approx \vec p_\nu$, is
imparted in each  neutrino collision. Due
 to the bulk velocity of 
the ``neutrino wind'' on Earth, $- \vec{\beta}_{earth}$, 
 there is a preferred
 direction for  $\Delta\vec p$, thus 
$\left<\Delta \vec p\right> \approx -\vec \beta_{earth}
 \left(3T_\nu/c\right)$ (or $ \approx -\vec \beta_{earth}~c~m_\nu $
 for clustered neutrinos).

The resulting accelerations for relativistic (Dirac or Majorana)
 neutrinos are~\cite{dgn}
\begin{equation}
a^D_R \simeq 2\times10^{-33}  {\rm cm}/{ \rm sec^2}~f (\rho/10)~.
 \label{aDR}
\end{equation}
For non-clustered non-relativistic Dirac neutrinos
 (i.e for most relic neutrinos), 
the cross-sections  are~\cite{dgn}
\begin{equation}
a^D_{NC-NR}\simeq 3\times10^{-27}  {\rm cm}/{ \rm sec^2}
 f \left(m_\nu/ { \rm eV}\right)^2~,
 (\rho/10)~,
 \label{aDNCNR}
\end{equation}
and for clustered non-relativistic Dirac neutrinos, 
the cross sections are~\cite{dgn}  
\begin{equation}
a^D_{C-NR} \simeq 10^{-26} {\rm cm}/{ \rm sec^2} f (\rho/10)~. 
 \label{aDCNR}
\end{equation}
 In these equations the factor $f$ accounts for the possible enhancement 
 due to clustering (as well as, possibly, a large
 lepton asymmetry), $1 \leq f \equiv (n_\nu +n_{\nu^c}) / 100$cm$^{-3} < 10$.

As mentioned above,  non-relativistic Majorana
 neutrinos have only a $\gamma_\mu \gamma_5$
coupling. The coherent interactions due to the
 static limit of this coupling are, however,
 suppressed by $\beta_\nu$, the ratio of the ``small"
 and ``large" components of the spinor.  Recall that for 
non-relativistic spinors the lower components are 
``smaller'' than the upper components 
by a factor of $\beta$.  Thus the 
analog of Eqs. (\ref{aDNCNR}) and (\ref{aDCNR}) for
non-relativistic Majorana neutrinos is suppressed by an extra factor of 
$\beta_\nu^2\approx 10^{-6}$.

\section*{6. Prospects for laboratory searches:  effects linear in $G_F$}

Coherent interations of a low energy neutral particle,
 with  a medium in which the interatomic spacing is
 much smaller than the deBroglie particle wavelength
(recall $\lambda_\nu \simeq 2.4$mm),
change the particle momentum from $p$ to $p'$. Then,
 one can define an index of refraction $n= p' /p$, and $n-1 \sim G_F$.
However, early proposals to use ``neutrino optics", either
   total reflection~\cite{opher} 
 or refraction (or refraction in a superconducting surface
  which would induce a current)~\cite{lewis}  were incorrect.
Cabibbo and Maiani~\cite{cabibbo} and  Langacker, Leveille and 
Sheiman~\cite{langacker} in 1982, proved that
 the force due to linear momentum or energy exchange on a
 target immersed in a uniform neutrino field cancel to order
 $G_F$, in fact~\cite{cabibbo}
\begin{equation}
\vec{F} = - \frac{\Delta \vec p_\nu}{\Delta t} \simeq G_F \int d^3x \rho_A (x)
\vec{\nabla} n_\nu(x)~,
 \label{noF}
\end{equation}
 where $\rho_A$ is the atomic number density of the target,
 and $\vec{\nabla} n_\nu(x)$ is the
 gradient of the local neutrino number density. This gradient is zero (since 
 $n_\nu$  due to gravitational effects is uniform at the scale
 of possible detectors), except 
for scattered waves due to weak interactions, which are of order
 $ G_F$, and thus lead to forces $F \simeq G_F^2$. In fact, 
Smith and  Lewin~\cite{smith}  proposed in 1983 to
 generate large artificial  neutrino density
distortions to induce a neutrino density gradient and 
thus a force, but there is no known way to do this.

There is  only  one possible mechanical effect of order $G_F$, 
proposed by  L. Stodolsky~\cite{stodolski} in 1974:
a torque of order $G_F$ can arise if both
 the target (e.g. consisting of magnetized iron) and the 
$\nu$-background have a polarization.
The $\nu$-background must have a non-zero  net flux of
 weak-interactions-charge  (i.e. of neutrinos minus antineutrinos) to
 produce a net torque on polarized electrons. 
Since the Earth is moving with respect to the C$\nu$B,
there is a net flux 
 of particles reaching us, with 
$\left<{\vec{\beta_\nu}}\right> = - \vec{\beta}_{earth}$. Thus  we only need a 
  lepton asymmetry in those particles to have a net flux of weak-charge 
$\sim - \vec\beta_{earth}(n_\nu - n_{\bar\nu})$
  
The  Stodolsky effect consists of an energy split of the two spin states of 
non-relativistic electrons in the C$\nu$B.   This energy split is
proportional to the difference between the densities of neutrinos and 
antineutrinos in the  neutrino background 
for Dirac neutrinos and relativistic Majorana, and  proportional
 to the net helicity of the
background for  non-relativistic Majorana neutrinos, as we will now see. The
distinction of Dirac or Majorana neutrinos  is important only for
 non-relativistic neutrinos.

The Hamiltonian density of the neutrino-electron interaction is
\begin{equation}
{\cal{H}}(x) = 
\frac{G_F}{\sqrt{2}}~ \left(\bar e\gamma^\mu(g_V - g_A \gamma_5)e\right)
\left[\bar\nu \gamma_\mu (1 - \gamma_5)\nu\right]~.
 \label{H}
\end{equation}
 In the non-relativistic limit the electron current
 $(\bar{e} \gamma^\mu \gamma_5 e)$ yields a 
factor  $\vec{\sigma}_e \cdot \vec\beta_e, \vec\sigma_e$
 (from the time and space components
of the current). For Dirac neutrinos (for which
 $\nu \neq \bar\nu$), the $\gamma_\mu $ term dominates,
 and
$ \left[ \bar\nu \gamma_\mu \nu \right] $ yields
 $ \bar\nu 1 \nu, ...\sim (n_\nu - n_{\bar\nu})$, 
 which is non zero if there is a net Lepton number
 in the C$\nu$B.  The effect, originally derived by
 Stodolski~\cite{stodolski} only for Dirac neutrinos,
 is proportional to the product 
$ \vec{\sigma}_e \cdot \vec\beta_e (n_\nu - n_{\bar\nu})$.
For Majorana neutrinos (for which $\nu = \bar\nu$), only the
 $\gamma_\mu \gamma_5$ term remains, and
 $\left[\bar\nu \gamma_\mu \gamma_5 \nu\right]$ 
 yield in the non-relativistic limit
$\bar\nu (\vec\sigma_\nu \cdot \vec\beta_\nu, \vec\sigma_\nu) \nu \sim 
(n_{\nu_\ell} - n_{\nu_r})$, which is non-zero 
only if there is a net helicity in the C$\nu$B~\cite{dgn}.

Here we call left (right)
chirality eigenstates $\nu_L(\nu_R),$ and 
left (right) helicity eigenstates
$\nu_{\ell}$ $(\nu_r)$.
The general expression for the energy of one electron
 in the C$\nu$B, in the electron-rest frame, to first
 order in $\beta_{earth}$, both for Dirac and Majorana
 neutrinos were first derived in Ref.\cite{dgn}. 
Let us see the most relevant particular cases, 
starting from early times, before
the decoupling of neutrinos.

Since neutrinos are lighter than about 1 eV, they
were relativistic at decoupling $(T_{\rm dec} \geq O({\rm MeV})$). 
Relativistic neutrinos are only in left-handed chirality states (and
anti- neutrinos only in right-handed chirality states). These are  
the only states produced
by weak interactions.   For relativistic neutrinos chirality and helicity 
coincide (up to mixing terms of order
 $m_{\nu}/ E_\nu \simeq m_{\nu}/T_\nu$). Thus,
at decoupling, neutrinos $\nu_L$ were
only in left-handed helicity states, and antineutrinos $\nu_R^c$ (or
$\nu_R$ in the case of Majorana neutrinos) in right-handed ones. 
In this case, the term in the Hamiltonian of one electron in the
C$\nu$B linear in the spin of the electron $\vec s_e$, is
\begin{equation}
H^D_R =  H^M_R= -\sqrt{2}G_F g_A ~2\vec 
s_e\cdot\vec\beta_{earth}(n_{\nu_L} -n_{\nu^c_R})~.
 \label{H1}
\end{equation}

Helicity is an eigenstate of
propagation and, therefore, it does not 
change while neutrinos propagate freely,
even if they become non-relativistic.
Recall that
two of the neutrinos mass eigentates are non-relativistic at present. 
For Majorana neutrinos chirality acts
as lepton number, so we are calling ``neutrinos" those particles produced
at $T>T_{dec}$ as $\nu_L$, and ``anti-neutrinos" those produced as $\nu_R$.
Thus, neglecting intervening interactions, non-relativistic
 background neutrinos 
are in left-handed helicity eigenstates (which consist of equal admixtures of
left- and right-handed chiralities) and anti-neutrinos are in right-handed
helicity eigenstates (which also consist of equal admixtures of
left- and right-handed chiralities). 
If the non-relativistic neutrinos are Dirac particles, only the left-handed
chirality states (right for anti-neutrinos) interact, since the  other 
chirality state is sterile, while if the
neutrinos are Majorana, both chirality states interact (the right-handed
``neutrino" state is the right-handed anti-neutrino). In the most 
favourable case of a large
 $\xi = \mu/T$,   for $m_\nu \leq 0.1 eV$, the term in the Hamiltonian
linear in the electron spin, for non-clustered
non-relativistic relic neutrinos (which are most of them), in the
 electron rest frame is
\begin{equation}
H^D_{NC-NR}\simeq \frac{1}{2} H^M_{NC-NR}
\simeq 0.85 \langle|\vec\beta_\nu|\rangle^{-1}
 H^D_R \leq \frac{7}{\sqrt{\xi}} H^D_R~, 
 \label{H2}
\end{equation}
where $\langle|\vec\beta_\nu|\rangle$=$\sqrt{6 \xi T_\nu/ 5 m}
  \simeq \sqrt{ \xi 1.7~10^{-4} {\rm eV}/m}$ is the characteristic 
velocity (the average of the
  velocity modulus) of relic neutrinos in the
  C$\nu$B rest frame. 
  The dominant term shown here, comes from the space part of the 
neutrino current ($\vec\sigma_\nu$ in the helicity base is
 $ h \hat\beta_\nu \sim \langle|\vec\beta_\nu|\rangle^{-1}$,
 where $\hat\beta_\nu$
  is the unit vector of the neutrino velocity in the relic
 neutrino rest frame). Notice that there is no $\beta$ factor
  penalty for Majorana neutrinos.

Slow enough non-relativistic neutrinos eventually fall into
 gravitational 
potential wells, become bound and,
 after a characteristic orbital time, their helicities 
become well mixed up, since momenta are reversed and  spins are not.  
Thus, gravitationally bound relic neutrinos  have well mixed helicities, no
net helicity remains. For these  clustered non-relativistc
 neutrinos, the Stodolski effect cancels for Majorana neutrinos,
but only changes by a factor for Dirac neutrinos.
\begin{equation}
H^D_{C-NR}= \frac{1}{2}H^D_R,  ~~~~~  H^M_{C-NR}= 0~.
 \label{H3}
\end{equation}
As we argued above, most relic neutrinos however are not gravitationally
 bound at present, because  they are too light.

The Stodolski effect requires a Lepton asymmetry
$n_\nu - n_{\nu^c} =  f 100$ cm$^{-3}$ (where the maximum possible
overdensity factor  is $f(\xi)_{max}=0.1-10$)
to obtain an energy difference, $\Delta E$, between the two helicity
 states of an electron in the direction of the bulk velocity of
 the neutrino background, $<\vec\beta_\nu>=-\vec\beta_{earth}$.
In the case of Dirac or relativistic Majorana neutrinos of
 density $n_\nu$, with a very 
large lepton asymmetry favouring, say, neutrinos $\nu_L$ (so that 
$n_{\nu_L}=n_{\nu_\ell}=  n_\nu$) we have
\begin{equation}
\Delta E \simeq  f 2 \sqrt{2} G_Fg_A~ |\vec\beta_{earth}|n_\nu
 \label{deltaE}~.
\end{equation}
This is the equation we will use to estimate the maximum possible
 accelerations due to this effect.
We should recall however that $\Delta E^M_{NC-NR} \leq 8 \Delta E$
 and that $\Delta E^M_{C-NR} = 0$.

The energy difference $\Delta E$  implies a torque 
of magnitude $\Delta E/\pi$  applied on the spin of the electron. 
 Since the spin is
``frozen" in a magnetized macroscopic piece of material with $N$
 polarized
electrons, the total torque applied to the piece has a magnitude 
 $\tau = N\Delta E/\pi$.
Given a linear dimension $R$ and mass $M$ of the macroscopic object, its
moment of inertia is parametrized as $I = MR^2/\gamma$, where $\gamma$ is a 
geometrical factor.  In the typical case of 
one polarized electron per atom in a material 
of atomic number $A$, the number $N$ above is 
$N = (M/gr) N_{\rm AV}/A$ (using cgs units), 
where $N_{\rm AV}$ is Avogadro's number. 
Thus, the effect we are considering would produce an
angular acceleration of order $\alpha = \tau/I$ and 
a linear acceleration of order $a_{G_F} = R\alpha$ in the magnet given by
\begin{equation}
a_{G_F} \simeq 10^{-27} \frac{\rm cm}{{\rm sec}^2} f 
\cdot\left(\frac{\gamma}{10}\right)
\left(\frac{100}{A}\right)\left(\frac 
{\rm cm}{R}\right)\left(\frac{\beta_{earth}}{10^{-3}}\right)~.
 \label{aGF}
\end{equation}
The local density enhancement factor due to a net
 lepton number or clustering
is $0 \leq f \equiv (n_\nu -n_{\nu^c}) / 100$cm$^{-3} < 100$.

We should compare the accelerations mentioned  in Eqs. (10), (11) 
and (18) above with 
 the smallest measurable acceleration at present,
 with ``Cavendish" type torsion balances, which is about 
$10^{-12}$ cm/sec$^2$\cite{minacc}. This is
about 15 orders of magnitude larger. As an example of an
 attempt to produce a relic neutrino detector,  we can
 mention a particular design  of a torsion oscillator 
proposed by C. Hagmann~\cite{hagmann}
 in 1999. In this proposed detector,
 the uncertainty principle gives a minimum measurable acceleration
(QL stands for quantum limit),
 \begin{equation}
 a_{\rm QL} = 5\times 10^{-24}{\rm cm/s^2}(10{\rm kg}/m)^{1/2}
({\rm 1 day}/\tau_0)^{1/2}(10^6 {\rm s}/\tau)
 \label{hami}
\end{equation}
 where $\tau_0$ is the oscillation period, and $\tau$ measurement
 time. This acceleration
 is larger by a factor (10$^3$-10$^4$)$f^{-1}$ than the largest
 possible accelerations mentioned
 above. We  would need a detector with a lower quantum limit,
and that can operate at that limit. To beat the quantum limit
P. Smith proposed  to  go to tons of mass target~\cite{smith, smith87} or to
 the opposite, to sub-micron granules whose single displacements
 can be measured (which involves nanotechnology 
not yet in place)~\cite{smith91}.

Still there are other difficulties.
Seismic and gravitational variations are a problem. For
 example, the gravity gradient due to the Moon produces a varying
 torque  about $10^{10}$ larger 
 than the relic neutrino force. A possible solution  would 
be a concentric balance~\cite{smith91}.
 Solar neutrinos would produce equal or larger accelerations
(and possibly dark matter particles could  too)~\cite{dgn}.
So directionality would be needed to separate a signal from
relic neutrinos  (Smith and Lewin~\cite{smith87}  
 proposed using laminated materials).

The evident conclusion is that  laboratory  effects of the 
C$\nu$B are still far from observability, therefore
awaiting  future technology.

\section*{6. Prospects for Astrophysical Searches}

Only at the Z-resonance the cross section of astrophysical
 neutrinos  with the C$\nu$B is large enough
to reveal its existence~\cite{weiler1}.  A simple argument 
is that the cross section at the Z-resonance,  
 $\sigma_{annih}(E^{\rm res}) \simeq 4\times 10^{-32}$cm$^2$, yields
 a mean free path for Ultra-High Energy
(UHE) neutrinos larger but not by much than the Hubble distance,
$\ell_{\rm Hubble}$,
 the size of the visible Universe,
$\ell_{m.f.p.} \simeq 35 \times \ell_{\rm Hubble}$.  Thus  the
 probability of interaction is non-negligible, about $ 0.03$.
 Otherwise, for interactions outside the Z-resonance, the
  Universe is transparent to UHE$\nu$.

If  an intense enough flux of UHE$\nu$ would exist the
 resonant annihilation with the C$\nu$B
would leave an absorption dip at 
\begin{equation}
E_{\nu_i}^{\rm res} = \frac{m_Z^2}{2\,m_{\nu_i}} = 
4.2\times 10^{22}\ {\rm eV}  \left( \frac{0.1\ {\rm eV}}{m_{\nu_i}}\right)~.
 \label{Eres}
\end{equation}
A recent examination~\cite{eberle} of the possibility of
 relic neutrino absorption spectroscopy through this mechanism
shows that, if intense  enough sources of UHE$\nu$ at energies
 $E_\nu \geq 10^{22}$ eV and above
would exist, the different masses of relic neutrinos could
 produce separated absorption dips. 
Experiment such as ANITA, Auger, EUSO, OWL, RICE, and SalSA
 able to detect such UHE neutrinos  are expected to be available
 in the near future~\cite{ANITAetc}. The problem with this idea,
 is that the
 only sources proposed to produce such UHE$\nu$ fluxes are topological
deffects. Even Active Galactic Nuclei are not thought to be
 able to produce neutrinos with energies above $10^{21}$ eV.

Another signature of the annihilation of UHE$\nu$ with relic
 neutrinos at the Z-resonance, is the
emission of UHE  p, n, $\gamma$ and $\nu$, in what T. Weiler
 called ``Z-bursts". In fact,
Z-bursts were proposed~\cite{weiler2} as the possible origin of UHE
 Cosmic Rays above the ``Greisen-Zatsepin Kuzmin cutoff" at
 $E_{GZK} \simeq 5\times 10^{19}$ eV observed, 
for example, by AGASA~\cite{AGASA}.
This mechanism is now considered unlikely to be correct  (and if
 it is correct, it would imply a relic neutrino mass 
$m_\nu \simeq 0.3$ eV~\cite{gvw}). If they do not produce the UHECR,
Z-bursts are subdominant, and thus not a good
 signal to detect the C$\nu$B.

\section*{7. Conclusions}

In the past, neutrinos were thought to be sufficiently
 massive ($m_\nu \ge 20$ eV)
 to cluster in our galaxy and make up the local dark matter
  halo. If so,
they could have had large local overdensities, even 
 $f \simeq  10^7$.
Also chemical potentials had  a much larger upper bound until a few years ago, 
$\xi \leq 6.9$~\cite{kang}. All this made for much more
optimistic predictions of the laboratory effects of relic neutrinos. 

Now, we know that relic neutrinos have sub-eV masses, the
 enhancement factors $f$ due to clustering could 
only be at most of $O(1)$,  and the maximum allowed 
chemical potential is likely $\xi \leq 0.1$.
We know also, that at least two of the three active 
neutrino mass eigenstates have masses
$m_{\nu_{2,3}} > 10^{-2}$~eV, thus most relic neutrinos 
 are non-relativistic and  non-clustered, and 
have an average velocity with respect to the Earth
 similar to that of the CMB, $\beta_\nu \simeq 10^{-3}$.
 These values lead to estimations of the
the effect of the ``cosmic neutrino wind" on macroscopic targets
 that are smaller than those made even a few
years ago. 

There are for sure forces on macroscopic targets of order $G_F^2$,
 and possibly 
torques of order $G_F$. $O(G_F^2)$ forces are largest when
 due to coherent elastic scattering out of target grains
 of the size of the neutrino deBroglie length $\lambda_\nu\simeq 0.2$ cm.
 The largest effect is for  Dirac neutrinos  with mass as
 large as possible (i.e. close to eV), as can be seen
 in Eqs. (\ref{aDNCNR}) and (\ref{aDCNR}).  The only 
 $O(G_F)$ effect is a torque on a polarized
  target, if there is a net flux of weak charge
due to the C$\nu$B. It requires a lepton asymmetry in the relic neutrinos.
The effect is largest for non-relativistic non-clustered 
 Majorana or Dirac neutrinos, for which the acceleration could be
 up to one order of magnitude larger than in Eq. (\ref{aGF}).
 These accelerations are tiny, and there are difficult backgrounds.
 New experimental ideas are needed to reach the required sensitivity.

With respect to the prospects for astrophysical searches,
Z-pole absorption dips in UHE$\nu$ flux
could reveal  relic neutrinos and their masses. The problem with this idea
is the existence of sources of UHE$\nu$ with $E_\nu \geq 10^{22}$ eV.
If these sources exist, 
this would be the most promising search mechanism in the foreseeable future,
using detectors such as ANITA, Auger, EUSO, OWL, RICE, and SalSA.

\bigskip\noindent
{\large \bf Acknowledgements}

This work is supported in part by the DOE Grant DE-FG03-91ER40662,
 Task C and NASA Grant NAG5-13399.

\end{document}